\documentclass[fleqn,10pt]{wlscirep}
\title{Experimental Trapped-ion Quantum Simulation of the Kibble-Zurek dynamics in momentum space}

\author[1,2,+]{Jin-Ming Cui}
\author[1,2,+]{Yun-Feng Huang}
\author[1,2]{Zhao Wang}
\author[1,2]{Dong-Yang Cao}
\author[1,2]{Jian Wang}
\author[1,2]{Wei-Min Lv}
\author[3,*]{Le Luo}
\author[4]{Adolfo del Campo}
\author[1,2,**]{Yong-Jian Han}
\author[1,2,***]{Chuan-Feng Li}
\author[1,2]{Guang-Can Guo}

\affil[1]{Key Laboratory of Quantum Information, University of Science and Technology of China,CAS,Hefei, 230026, People's Republic of China}
\affil[2]{Synergetic Innovation Center of Quantum Information and Quantum Physics,University of Science and Technology of China, Hefei, 230026, People's Republic of China}
\affil[3]{Trapped Atoms and Ions Laboratory, Department of Physics,Indiana University-Purdue University Indianapolis, Indianapolis, IN 46202-3273}
\affil[4]{Department of Physics, University of Massachusetts Boston,100 Morrissey Boulevard, Boston, MA 02125}

\affil[*]{leluo@iupui.edu}
\affil[**]{smhan@ustc.edu.cn}
\affil[***]{cfli@ustc.edu.cn}

\affil[+]{these authors contributed equally to this work}

\begin{abstract}
	The Kibble-Zurek mechanism is the paradigm to account for the nonadiabatic dynamics of a system across a continuous phase transition. Its study in the quantum regime is hindered by the requisite of ground state cooling. We report the experimental quantum simulation of critical dynamics in the transverse-field Ising model by a set of Landau-Zener crossings in pseudo-momentum space, that can be probed with high accuracy using a single trapped ion. We test the Kibble-Zurek mechanism in the quantum regime in the momentum space and find the measured scaling of excitations is in accordance with the theoretical prediction.
\end{abstract}
\begin{document}
	
	\flushbottom
	\maketitle
	%
	%
	\thispagestyle{empty}
	
	\section*{Introduction}
	
	Phase transitions are ubiquitous in physics, and account for the fundamental change of the state of a system when a control parameter is driven across a critical point. Continuous phase transitions are characterized by the divergence of the correlation length and relaxation time. As a result of the critical slowing down, the dynamics induced by a finite-time quench across the critical point is generally nonadiabatic and results in the formation of topological defects. The implications of the nonadiabatic nature of the critical dynamics during symmetry breaking were discussed by Kibble \cite{Kibble1976,Kibble1980} in a cosmological setting. Soon after, Zurek pointed out that condensed matter systems provide a test-bed for this scenario \cite{Zurek1985,Zurek1996}, and predicted that the density of the topological defects scales with the transition rate $1/\tau_{Q}$ as a universal power law of the form $n\propto\tau_{Q}^{-\beta}$, where $\beta>0$ is a function of the critical exponents of the phase transition and the dimensionality of the system. The so-called Kibble-Zurek mechanism (KZM) shows that the equilibrium critical exponents can be used to predict the density of defects formed in the non-equilibrium critical dynamics and constitutes a prominent paradigm in non-equilibrium physics \cite{Eisert2015}.
	
	The applicability of the KZM is expected to extend from low-temperature physics to cosmology. Experimental efforts in a variety of systems including liquid crystal \cite{Chuang1991,Bowick1994}, superfluid helium 3 \cite{Bauerle1996} and 4 \cite{Hendry1994}, low temperature Bose-Einstein condensates \cite{Weiler2008,Ferrari13NP}, trapped ions \cite{Delcampo2010,Ulm2013,Pyka2013},  multiferroic hexagonal magnanites \cite{Griffin12PRX} and colloidal systems \cite{Keim15PNAS} have verified aspects of the mechanism, while the scaling of excitation as a function of the quench rate across the quantum critical point remains elusive \cite{Zurek14IJoMPA}. The situation is even more uncertain in the quantum regime. Theoretical studies \cite{Zurek2005,Dziarmaga2005,Polkovnikov05PRB,Dziarmaga10AiP} and moderate experimental progress \cite{Chen2011PRL, Baumann2011PRL, Braun2015PNAS} support the tenets of the mechanism, but a direct test is still missing. Despite the development of cooling techniques in optical lattice \cite{Hamann1998} and trapped ion systems \cite{Diedrich1989}, the experimental study of the KZM across a quantum phase transition remains challenging. This is due to the fact that the system of interest should be cooled to its ground state, and that an accurate control of the parameter driving the transition over a large range of values is required. In addition, the resulting non-equilibrium state should be probed with high-efficiency. Regarding the choice of the system, the transverse field Ising model (TFIM) is the paradigmatic model to study quantum phase transitions \cite{Sachdec}, and constitutes an ideal test-bed to investigate the KZM in the quantum regime. However, simulation of this spin system in optical lattice can only be realized in some special situation \cite{Greiner2011}. By contrast, it has been realized in trapped ion systems \cite{Monroe2010,Blatt2012,Monroe2013,Monroe2014}, but with a limited system size preventing the test of the KZM.
	
	In this work we report the quantum simulation of the critical dynamics in the TFIM with a well controlled trapped ion in momentum space. We exploit the mapping of the TFIM critical dynamics to a set of independent Landau-Zener (LZ) crossings in pseudo-momentum space \cite{Dziarmaga2005}. Firstly, we study an essential ingredient of the KZM in our setup, the so-called adiabatic-impulse (AI) approximation, that successfully accounts for the dynamics of a single LZ crossing \cite{Damski2005,Damski2006}. By comparison to previous experiments \cite{Wang2014}, trapped ion qubit exhibits long coherence times\cite{Olmschenk2007}. After exploiting this fact, we study the universal scaling of excitation as a function of the quench rate across the quantum critical point of the TFIM. Our experimental results show that the measured density of the excitations is in accordance with the theoretical prediction of the KZM in the quantum regime.
	
	\section*{Results}
	
	\subsection*{Theory}
	In one dimension, the Hamiltonian of the TFIM with linear quench is defined by
	\begin{equation}
		H=-J\sum_{n}^{N}\left[g(t)\sigma_{n}^{x}+\sigma_{n}^{z}\sigma_{n+1}^{z}\right]=-J\sum_{n}^{N}\left[\Delta_{I} t\sigma_{n}^{x}+\sigma_{n}^{z}\sigma_{n+1}^{z}\right],\label{TFIM}
	\end{equation}
	where $N$ is the number of spins, $g(t)$ is the amplitude of the magnetic field, $\Delta_I$ is the quenching rate, $\sigma_{x}$ and $\sigma_{z}$ are Pauli matrices, and $J$ sets the energy scale of the system. In what follows we consider periodic boundary condition such that $\vec{\sigma}_{N+1}=\vec{\sigma}_{1}$ and denote by $|\rightarrow\rangle$ the eigenvector of $\sigma_{x}$ with eigenvalue $1$ and by $|\uparrow\rangle$ and $|\downarrow\rangle$ the eigenvectors of $\sigma_{z}$ with eigenvalues $1$ and $-1$, respectively. As the parameter $g(t)$ is changed from $+\infty$ to $0$, the Ising chain exhibits a quantum phase transition in which the ground state changes from a paramagnetic state, $|\rightarrow\rightarrow\cdots\rightarrow\rangle$ to a doubly degenerate ferromagnetic ground state, with all spins pointing up $|\uparrow\uparrow\cdots\uparrow\rangle$ or down $|\downarrow\downarrow\cdots\downarrow\rangle$. The critical point is located at $g_{c}=1$. When varying $g(t)$ across this value, adiabatic driving is not possible due to the closing of the gap between the ground state and the first excited state as $g(t)\rightarrow g_{c}$. As a result, the dynamics of the phase transition is intrinsically nonadiabatic and formation of kinks at the interface between different domains is expected. These excitations, quantum in nature, can be pictured as coherent superpositions of the form $|\downarrow\downarrow\cdots\downarrow\uparrow\uparrow\cdots\uparrow\downarrow\downarrow\cdots\downarrow\rangle$. Consider the preparation of the system in the paramagnetic ground state of the Hamiltonian with $g(t)\gg g_{c}=1$, followed by crossing of the phase transition by ramping down the magnetic field $g(t)$ to a final zero value. The time-dependence of $g(t)$ can be linearized in the proximity of the critical point as $g(t)=-t/\tau_{Q}$, where $\tau_{Q}$ sets the quench time scale in which the transition is crossed and the time of evolution varies from $t/\tau_Q\ll-1$ to 0. According to KZM, the density of the kinks after the transition scales as a universal power-law with the quench rate
	\begin{equation}
		n_{KZM}\approx\sqrt{\frac{\hbar}{2J\tau_{Q}}}.\label{est}
	\end{equation}
	
	Generally, the exact form of the density of the topological defects in a many-body system is difficult to determined. However, as pointed out by Dziarmaga \cite{Dziarmaga2005}, the density of kinks in the TFIM can be exactly computed. Using the Jordan-Wigner and Bogoliubov transformations, the TFIM can be mapped to a set of independent quasi-particles. The resulting density of kinks at the end of the quench $t=0$, can be measured via the expectation value of the operator
	\begin{equation}
		n_{ex}=\frac{1}{N}\sum_{n}^{N}\frac{1}{2}\langle1-\sigma_{n}^{z}\sigma_{n+1}^{z}\rangle=\frac{1}{N}\sum_{k}^{N}\langle\gamma_{k}^{\dagger}\gamma_{k}\rangle,\label{kKZM}
	\end{equation}
	where $\gamma_{k}$ is the annihilation operator of the quasi-particle with pseudo-momentum $k$ at $t=0$ and the pseudo-momentum $k$ take $N$ different values $\{\pm\frac{1}{2}\frac{2\pi}{Na}$,$\dots$,$\pm(\frac{N}{2}-\frac{1}{2})\frac{2\pi}{Na}\}$, where $a$ is the lattice spacing. The operator $\gamma_{k}$ is related to the initial Pauli matrices by the Bogoliubov-de Gennes and Jordan-Wigner transformation \cite{Dziarmaga2005}. Therefore, different modes decouple in Fourier space and the density of the kinks at $t=0$ can be calculated as the sum $\langle\Psi(0)|\sum_{k}\gamma_{k}^{\dagger}\gamma_{k}|\Psi(0)\rangle$$/N$, where $\Psi(0)$ is the finally state of the quench at $t=0$. The dynamics of independent quasiparticles is described by the time-dependent Bogoliubov-de Gennes equation, whose solution determines the excitation probability in each mode, $p_{k}=\langle\Psi(0)|\gamma_{k}^{\dagger}\gamma_{k}|\Psi(0)\rangle$ \cite{Dziarmaga2005}. The dynamics in each mode can be cast in the form of a Landau-Zener (LZ) problem,
	\begin{equation}
	i\hbar\frac{d}{d\tau_k}\left[\begin{array}{c}
	\nu_{k}\\
	\mu_{k}
	\end{array}\right]=\frac{1}{2}J\left[\begin{array}{cc}
	\Delta_{k}\tau_k & 1\\
	1 & -\Delta_{k}\tau_k
	\end{array}\right]\left[\begin{array}{c}
	\nu_{k}\\
	\mu_{k}
	\end{array}\right],
	\end{equation}
	where $\mu_{k}$ and $\nu_{k}$ are time-dependent Bogoliubov parameters for pseudo-momentum $k$, $\tau_k=4\tau_{Q}\sin(ka)(t/\tau_{Q}+\cos(ka))$, $\Delta_{k}^{-1}=4\tau_{Q}\sin^{2}(ka)$ and $\tau_Q=1/\Delta_I$. Comparing with the standard LZ model
	$H_{LZ}(t)=\frac{1}{2}J\left(\Delta t\sigma_{z}+\sigma_{x}\right)$,
	the parameters $\tau_k$ and $\Delta_{k}$ play the role of $t$ and $\Delta$, respectively. As a result, each probability $p_{k}$ can be found by carefully controlling the parameters $\Delta_{k}$, varying $\tau$ from $-\infty$ to $\tau_{k,f}=2\tau_{Q}\sin(2ka)$ (corresponding to $g(t)=0$), and measuring the probability of the excited state in the final state. Using the LZ formula for $p_{k}$ \cite{Zener1932}, the density of the kinks can be calculated as \cite{Dziarmaga2005}:
	\begin{equation}
		n_{ex}=\frac{1}{2\pi}\sqrt{\frac{\hbar}{2J\tau_{Q}}}.\label{Dkink}
	\end{equation}
	This agrees with the universal power-law scaling predicted by the KZM (\ref{est}), and further shows that the prefactor is overestimated by KZM.

	To experimentally study the density of the kinks in the TFIM resulting from a sweep through the critical point, an accurate control of the LZ model is necessary over a large range of values of the control parameter. In \cite{Xu2014}, a special situation of the LZ supporting the AI approximation has been demonstrated in linear optics. However, that setup could not be used to implement a LZ crossing for the long evolution times required to study critical dynamics. By contrast, a trapped ion allows for a long coherence time and a high fidelity initializing, manipulating, and detecting the ion quantum state \cite{Lucas2014}, making it an ideal platform to complete the full evolution in the LZ model and investigate the KZM in the TFIM.
	
	The LZ model is realized with a single $^{171}\mathrm{Yb^{+}}$ ion  confined in a Paul trap, consisting of six needles placed on the $x$-$z$ and $y$-$z$ planes, as shown in Fig. 1A. The hyperfine clock transition in the ground state $S_{1/2}$ manifold is chosen to realize the qubit, with energy levels denoted by $\left|0\right\rangle \equiv\left|F=0,\, m_{F}=0\right\rangle $ and $\left|1\right\rangle \equiv\left|F=1,\, m_{F}=0\right\rangle $  \cite{Olmschenk2007}. In zero static magnetic field, the splitting between $\left|0\right\rangle $ and $\left|1\right\rangle $ is 12.642812 GHz. We applied a static magnetic field of 4.66 G to define the quantization axis, which changes the $\left|0\right\rangle $ to $\left|1\right\rangle $ resonance frequency to 12.642819 GHz and creates a 6.5 MHz Zeeman splittings for $\mathrm{^{2}S_{1/2},\, F=1}$. In order to manipulate the hyperfine qubit with high control, the coherent driving is implemented by a microwave with IQ modulation, see the scheme in Fig. 1C. A two channel Arbitrary Function Generator (AFG) creates signals around 2 MHz to modulate 3.0 GHz microwave from SG384, a 9.64 GHz microwave is mixed with the modulated 3.0 GHz signal to get an arbitrary microwave near 12.642 GHz, and then the signal is amplified to 2\,W and irradiated to the trapped ion by a horn antenna. We first use Doppler cooling and initialize the qubit into the $\left|0\right\rangle $ state by optical pumping as in \cite{Olmschenk2007}, then send the microwave sequence to the ion, and finally measure the population of the bright state $\left|1\right\rangle $ by fluorescence detection scheme \cite{Olmschenk2007}.
	
	\subsection*{LZ dynamics and the AI approximation}
	For the quantum simulation of the KZM, we first characterize the LZ dynamics. According to Damski \cite{Damski2005} a LZ crossing can be well described by the adiabatic-impulse (AI) approximation which plays the central role in the KZM.

	Consider the time-dependent Hamiltonian $H=\frac{1}{2}J(\Delta t\sigma_z+\sigma_x)=\frac{1}{2}\hbar(vt\sigma_z+\Omega_0\sigma_x)$, where $v$ is the energy detuning rate and $\Omega_0$ is the Rabi frequency of the system when detuning $vt=0$. The AI approximation splits the time evolution during a LZ crossing into three sequential stages during which the dynamics is adiabatic, effectively frozen, and adiabatic again. The three regions can be separated by the freeze-out time scale $\pm\hat{t}$, at which the equilibrium relaxation time $\tau=\hbar/{\rm gap}(t)$ set by the inverse of the instantaneous gap matches the time measured from the avoided crossing, e.g. $\tau(\hat{t})=\hat{t}$. The impulse region extends over the interval ($-\hat{t}$,$\hat{t}$), when the system can not follow the external quench and is effectively frozen. Considering the difference between the LZ and second order phase transition near $t=0$ (the gap closes at the critical point),  the energy gap finite as $\hbar/ \mathrm{gap(t)}=1/{\sqrt{\Omega_0^{2}+(v t)^{2}}}=\tau_0/{\sqrt{1+(t/\tau_{Q})^{2}}}$  (where $\tau_0=1/\Omega_0$ and $\tau_{Q}=1/ \Delta$), the equation determining the freeze-out time should be modified in a LZ crossing with a linear parameter $\alpha$ as \cite{Damski2006}: $\hbar/ \mathrm{gap}(\hat{t})=\alpha\hat{t}$.
	
	Using the AI approximation, the probability of the excited state in LZ can be predicted as the overlap between the instantaneous ground state  at $-\hat{t}$ and the excited state at $+\hat{t}$\cite{Damski2005}. For the generic scheme A, the initial state is prepared at the ground state of the system which is far away from the critical point. The probability of the excited state at the end time $t_{f}\rightarrow\infty$ is then given by
	$\mathcal{D}={2}/{\mathcal{P}(x_{\alpha})}$,
	where $\mathcal{P}(x_{\alpha})=x_{\alpha}^{2}+x_{\alpha}\sqrt{x_{\alpha}^{2}+4}+2$, $x_{\alpha}=\alpha\tau_{Q}/\tau_0$.
	
	Experimentally, after Doppler cooling and initializing the ion state into $\left|0\right\rangle $, a microwave sequence is applied to the ion. For case A, the microwave is a phase continuous frequency sweep waveform near the spin transition. The time dependence of a typical LZ transition probability to $\left|1\right\rangle $ is shown in Fig. 2B, where $t=0$ is set by the microwave frequency resonating with the clock transition.  $\Omega_{0}=18.3$ KHz, $v=2.0$ GHz/s, and the time evolution starts from $t_{i}=-500\,\mu s$.
	We measured the probability in the excited state for different values of the quench time $\tau_{Q}$, by changing the frequency sweeping rate. Data and fitting results are plotted in Fig. 2C. The fitted scaling parameter $\alpha=1.58\pm0.01$ is in agreement with the theoretical value $\pi/2$ \cite{Damski2006}.
	
	In the scheme B, the initial state is prepared in the ground state of the system at anticrossing center, $t_{i}=0$. As $t_{f}\rightarrow\infty$,
	$\mathcal{D}=\frac{1}{2}\left(1-\sqrt{1-2/\mathcal{P}(x_{\alpha})}\right)$.
	A typical transition probability as a function of time is shown in Fig. 2F. In order to prepare the initial state in the ground eigenstate, which is $\frac{\sqrt{2}}{2}(\left|0\right\rangle +\left|1\right\rangle )$, we implement a $R_{Y}(\frac{\pi}{2})$ qubit rotation microwave pulse before the microwave frequency sweep. The fitted scaling parameter to the measurement data for different quench times, $\alpha=0.78\pm0.03$, is in accordance with the theoretical prediction $\pi/4$ \cite{Damski2006}.
	
	\subsection*{Quantum simulation of the KZM in TFIM}
	We have shown that a trapped ion can be used to simulate the LZ process with high accuracy. The exquisite control of this system renders the study of the KZM in TFIM feasible. To this end, we implement the following experimental scheme with a trapped ion:
	\begin{enumerate}
		\item The Rabi frequency of the system $\Omega_{0}$,  that corresponds to the energy scale  $J=\hbar\Omega_{0}$ in the simulated Ising model, is measured and kept stable during the whole experiment. For a given quench time $\tau_{Q}$, the experimental values of $\tau_{k,f}$ and $\Delta_{k}$ required to control the microwave at different k-modes are found via the equations $\tau_{k,f}=2\tau_{Q}\sin(2ka)$ and $\Delta_{k}^{-1}=4\tau_{Q}\sin^{2}(ka)$.
		\item Waveforms of first momentum parameter ($k_{1}$) are computer-generated and set to AFG channels; the experiment is run repeatedly to estimate the mean density of excitations, according to the experiment control process shown in Fig. 3A. Photon counts from the bright state are recorded after $\left|1\right\rangle$ is prepared in the reference pulse, which is used to normalize photon counts of the excitation pulse, to get the excitation probability.
		\item
		This procedure is repeated to get the excitation probability in each mode, $p_{k}$.
	\end{enumerate}
	The $k$-mode control sequence consists of a wave with sweeping rate of $v_k=\Delta_{k}\Omega_0$ and an operation pulse $R_{X}(\pi)R_{Y}(\theta_{k})$. The Rabi frequency of the system is typically 20 KHz; we start to sweep microwave with 2 MHz detuning, so that the initial condition is located far away from anticrossing point. The $R_{X}(\pi)R_{Y}(\theta_{k})$ operation pulse transforms the excited state of the Hamiltonian at $\tau_{k,f}$ to the ion bright state $\left|1\right\rangle $, with a rotation angle $\theta_{k}=\Delta_{k}\cdot\tau_{k,f}$.

	Implementing this scheme, the excitation probability in the broken-symmetry phase of a quantum Ising chain of 50 spins is measured, see Fig. 3B. Black, red and blue dots denote data measured with relative quench time at 1.85, 0.85 and 0.35, respectively. The solid lines correspond to the numerical simulations. Applying equation $P=\sum p_{k}/N$, kink densities for different quench times can be obtained. Fig. 3C depicts the variation of the kink density with $\hbar(J\tau_{Q})^{-1}$. The scaling behavior becomes apparent in a double logarithmic plot, and a fit to the data leads to $n_{ex}\sim\tau_{Q}^{-0.59\pm0.03}$ that accords with the power-law predicted by the KZM across the quantum phase transition.
	
	\section*{Discussion}
	
	We have demonstrated that a well-controlled trapped ion offers a quantum platform to study the KZM in TFIM in pseudo-momentum space by exploiting the mapping of the critical dynamics to a set of independent LZ processes. Our experiment test the validity of the AI approximation in the LZ and the quantum KZM in the TFIM. The power-law exponent measured in our experiment is $0.59\pm0.03$ matches the numerical simulation in \cite{Zurek2005} which reported the value $0.58$ for the quantum Ising chain. In addition, the prefactor of the density of excitations in our experimental results is $0.136\pm0.004$ which agrees with the $0.16$ estimated from numerical simulations  in \cite{Zurek2005}. As a result, our experimental quantum simulation reveals the signature of universality in the quantum critical dynamics, the main prediction of the KZM.
	We notice that our scheme can be readily applied to test a variety of protocols, including the critical dynamics under inhomogeneous driving \cite{Rams10NJoP}, nonlinear quenches \cite{Polkovnikov08PRL,Sinitsyn13PRL}, and the simulation of multiple-body interactions required for counterdiabatic driving and the suppression of the KZM across a quantum phase transition \cite{Zurek12PRL,Takahashi13,Damski14JoSMTaE}.  As a result, it constitutes an ideal experimental platform  for the quantum simulation of nonequilibrium many-body systems in momentum space. The limitations of our setup is that it can only simulate the KZM dynamics in momentum space and cannot probe topological defect formation in real space.
	
	\section*{Methods}
	
	\textbf{Ion trap setup. }
	The size of the needle trap  depends mainly on the distance between the two needles tips   near the trap center, which is set to 120 $\mu\mathrm{m} $ in the experiment. The  needle trap is installed in an ultrahigh vacuum below $10^{-11}$ torr, and a helical resonator provides the RF signal  with  frequency of 27\,MHz and  amplitude of 150\,V  to the trap, that sets the ion secular motion frequencies to 1.45 MHz, 1.66 MHz and 2.96 MHz. Ion fluorescence is collected by an objective lens with 0.4 numerical aperture (N.A), and detected by a photomultiplier tube (PMT). Total fluorescence detection efficiency is  $\eta_{total}=\eta_{lens}\cdot\eta_{trans}\cdot\eta_{detect}$$=4\%\times90\%\times30\%$=$1.1\%$
	
	\noindent\textbf{Waveform of the driving microwave. }
	In principle, any required waveform of the microwave field can be generated by setting the waveforms of dual channel AFG for IQ modulation.  We consider the carrier microwave $B_{c}(t)=A\sin(\omega_{c}t)$, with amplitude $A$ and frequency $f_{c}=\omega_{c}/2\pi$. The waveforms generated by AFG for IQ modulation are $I(t)=\cos(\phi(t))$ and $Q(t)=\sin(\phi(t))$ respectively. After IQ modulation, the microwave field will be $B(t)=A\sin(\omega_{c}t+\phi(t))$, where the phase function $\phi(t)$ can be expressed in a section function for the microwave composed of a series of frequency sweeping waveforms and qubit rotation pulses. With a qubit resonance frequency $f_{0}=\omega_{0}/2\pi$, we sweep the microwave frequency at $\omega_{1}$ from $t=0$, with sweeping rate $v$ to $t=t_{1}$. Then the qubit is manipulated with $R_{y}$ in $(t_{1},\, t_{2})$ and $R_{x}$ in $(t_{2},\, t_{3})$. The expression  of $\phi(t)$ can be derived
	\begin{eqnarray}
		\phi(t)=\begin{cases}\frac{1}{2}v t^{2}+(\omega_{1}-\omega_{c})t & ,(0,\, t_{1})\\
		(\omega_{0}-\omega_{c})t+\phi_{1}+\frac{\pi}{2} & ,(t_{1},\, t_{2})\\
		(\omega_{0}-\omega_{c})t+\phi_{2} & ,(t_{2},t_{3})
		\end{cases},
	\end{eqnarray}
	where $\phi_{1}=\frac{1}{2}v t_{1}^{2}+(\omega_{1}-\omega_{c})t_{1}$, $\phi_{2}=\phi_{1}+(\omega_{0}-\omega_{c})(t_{2}-t_{1})$.
	
	In actual experiment, there is a residual carrier wave with small amplitude $\epsilon_{c}$, resulting from the imperfection of the IQ modulation $B_{c}(t)=\epsilon_{c}\sin(\omega_{c}t)$. While the ratio of amplitudes $\epsilon_{c}$ and $A$ is small, $\epsilon_{c}/A<0.01$, when the scheme works in resonance modulation $\omega_{0}=\omega_{c}$ the residual carrier wave can change the qubit state inducing an error over one percent after several Rabi flips. In order to reduce this error, we set $\omega_{c}$ such that $\omega_{0}-\omega_{c}=(2\pi)$2.0 MHz, which is largely detuned from resonance compared to the  Rabi frequency $\Omega_{0}=(2\pi)$20 KHz, and renders the residual wave drive negligible.
	

	\section*{Acknowledgements}
	
	We would like to thank Prof. Long-Sheng Ma for providing help on stabilizing lasers and technical suggestions.  This work was supported by the National Natural Science Foundation of China (grant nos. 11274289, 11325419, 11474267, 11404319, 61327901, 61225025 and 11474268), the Fundamental Research Funds for the Central Universities (grant nos. WK2470000018 and WK2030020019), the Strategic Priority Research Program (B) of the Chinese Academy of Sciences (Grant No. XDB01030300) and UMass Boston (project P20150000029279).
	
	\section*{Author contributions statement}
	
	J.M.C., Y.F.H., L.L. and C.F.Li designed the experiment. Y.J.H. and A.d.C. did the theoretical analysis. J.M.C., Y.F.H., Z.W., D.Y.C., J.W. and W.M.Lv performed the experiment. J.M.C. and Y.J.H. and A.d.C. wrote the manuscript. C.F.Li and G.C.G. supervised the project.  All authors discussed and contributed to the analysis of the experimental data.
	
	\section*{Additional information}
	\textbf{Competing financial interests:} The authors declare that they have no competing financial interests.
	
	\newpage
	
	\begin{figure}[ht]
		\includegraphics[width=0.95\columnwidth]{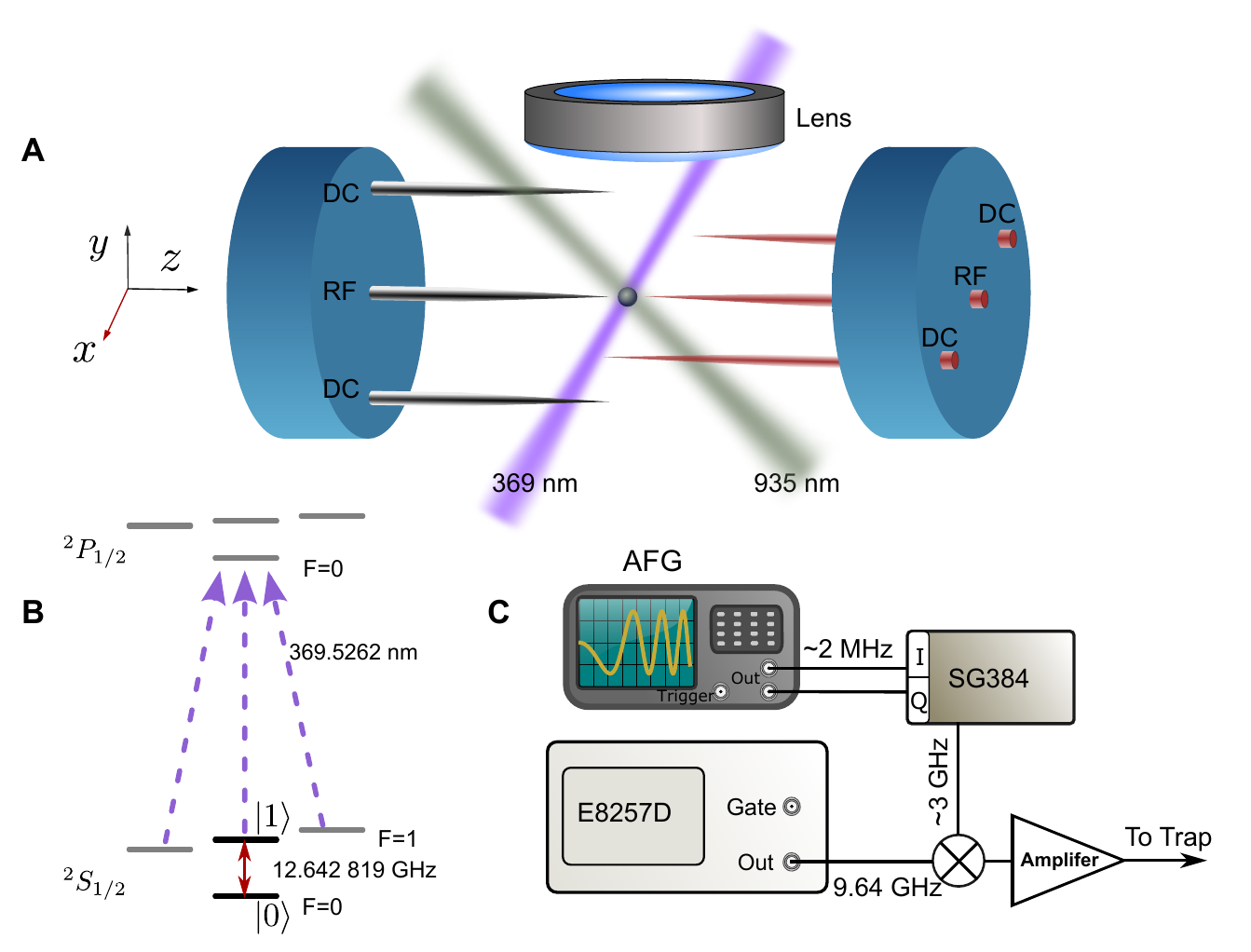}
		
		\caption {{\bf \label{fig:setup} Experimental setup using an ion-trap.}  ({\bf A}) Configuration of six needles of the trap used in experiment. ({\bf B}) $^{171}\mathrm{Yb^{+}}$ energy spectrum: the hyperfine energy levels ($\left|0\right\rangle $ and $\left|1\right\rangle $) of the ground state are used as the qubit. ({\bf C}) Microwave control scheme of the driving field around 12.642 GHz: the frequency is generated by mixing microwaves between 9.64 GHz and 3 GHz, while 3 GHz microwave is IQ modulated by a dual channel AFG around 2 MHz.}
	\end{figure}
	
	\begin{figure}[ht]
		\includegraphics[width=1.0\columnwidth]{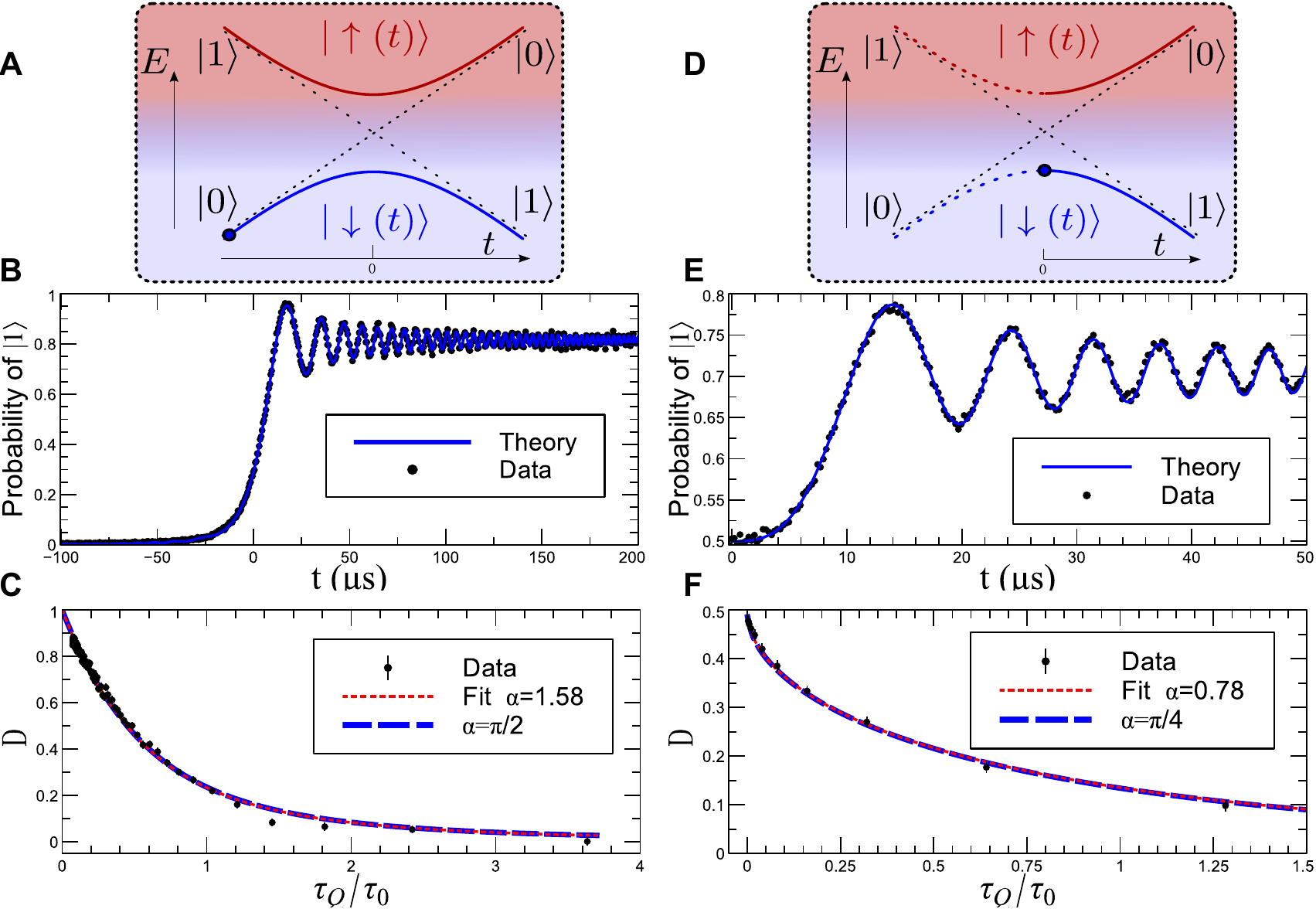}
	
		\caption {{\bf \label{fig:LZ} LZ transition and density of detects for different cases.}  ({\bf A}), ({\bf D}) schematic energy level and initial condition of two evolution  case, respectively. ({\bf B}) A typical plot of measured probability of $\left|1\right\rangle $ state of $^{171}\mathrm{Yb^{+}}$ during the LZ crossing with evolution time starting from $t=-500\,\mu$s for ({\bf A}), experimental parameter: $\Omega_{0}=18.3$ KHz, $v=2.0$ GHz/s; ({\bf E}) a typical plot of measured probability of $\left|1\right\rangle $ with starting time at $t=0$ with experimental parameters: $\Omega_{0}=13.8$ KHz, $v=5.0$ GHz/s, black dots: data, blue lines: numerical simulation. ({\bf C}), ({\bf F}) Density of defects for ({\bf A}), ({\bf D}), respectively, black dot: measured data, red dot line: the fitting line, blue dash line: theory curve in \cite{Damski2006}.}
	\end{figure}
	
	\begin{figure}[ht]
		\includegraphics[width=0.85\columnwidth]{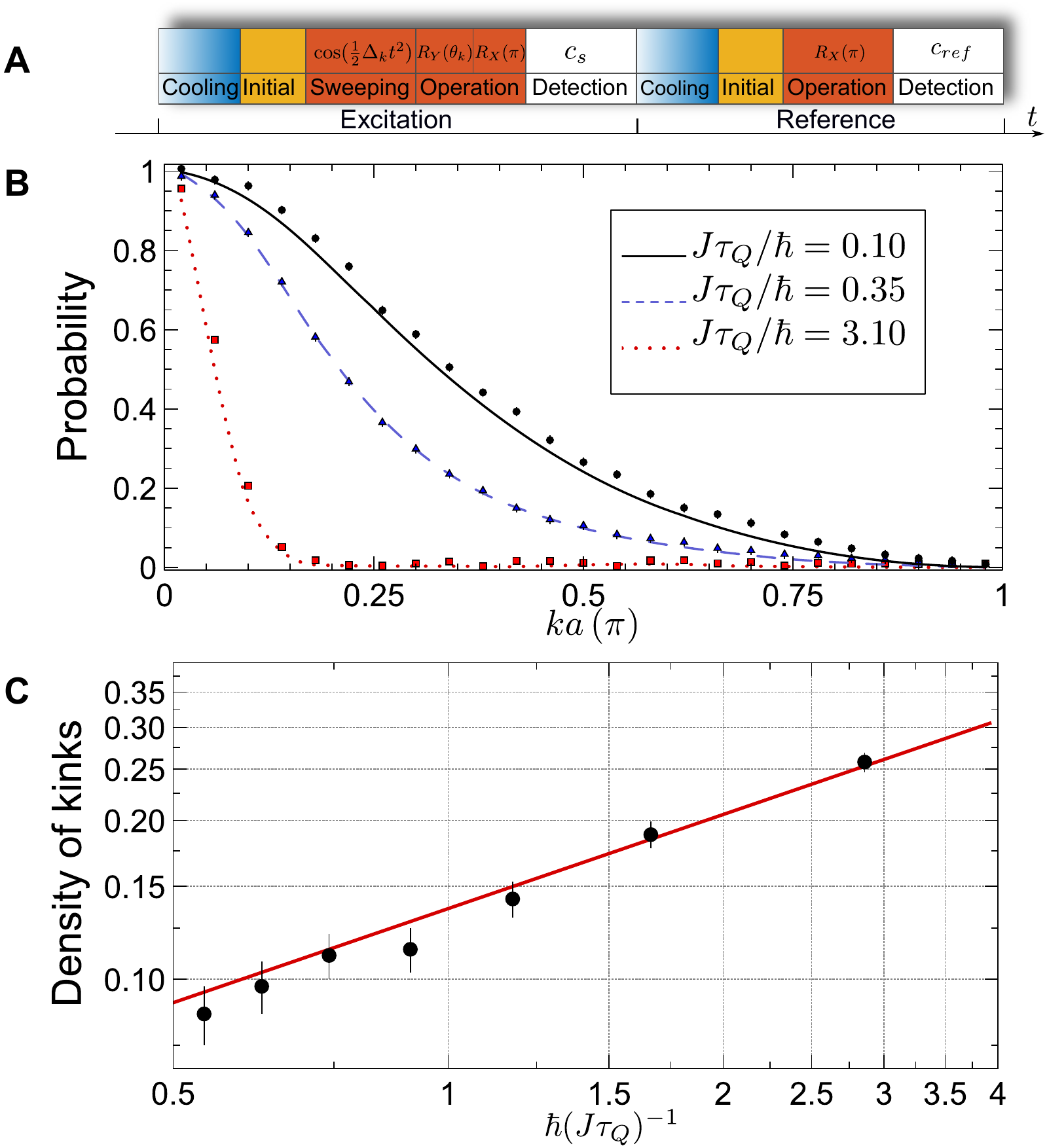}
	
		\caption { {\bf \label{fig:Ising} Quantum simulation of the critical dynamics in the TFIM with a series of uncoupled LZ crossings.}  ({\bf A}) Control process for measurement as a function of time; ({\bf B}) Excitation probabilities in each of the $k$ modes, black dots, red squares and blue triangles are data measured with $J\tau_{Q}/\hbar$ at 1.85, 0.85 and 0.35, respectively. Corresponding lines are numerical calculation; ({\bf C}) Density of kinks ($n_{ex}$) for different quench time, the fitted scaling parameter is $n_{ex}\sim\tau_{Q}^{-0.59\pm0.03}$.}
	\end{figure}
	
\end{document}